\documentclass[prd,nofootinbib,preprint,showpacs,showkeys,superscriptaddress]{revtex4}
\usepackage[english]{babel}
\usepackage{amsmath,amssymb}
\usepackage[dvips]{graphicx}
\usepackage{ifthen,array}
\usepackage[sort&compress]{natbib}
\usepackage{amsfonts}
\usepackage{bbm}
\usepackage{amssymb}
\newcommand{\Dms}  {\Delta m^2_\Sol}
\newcommand{\Dma}  {\Delta m^2_\Atm}
\DeclareMathAlphabet{\mathsc}{OT1}{cmr}{m}{sc}

\newcommand{\Sol}  {\mathsc{sol}}
\newcommand{\Atm}  {\mathsc{atm}}
\renewcommand{\baselinestretch}{1.3}
\allowdisplaybreaks

\def\10{$SO(10)$}
\def\21{SU(2) $\otimes$ U(1) }

\def\422{$SU(4) \otimes SU(2) \otimes SU(2)$}
\def\321{SU(3) $\otimes$ SU(2) $\otimes$ U(1)}

\def\lsim{\raise0.3ex\hbox{$\;<$\kern-0.75em\raise-1.1ex\hbox{$\sim\;$}}}
\def\gsim{\raise0.3ex\hbox{$\;>$\kern-0.75em\raise-1.1ex\hbox{$\sim\;$}}}

\newcommand{\AddrAHEP}{%
  AHEP Group, Institut de F\'{\i}sica Corpuscular --
  C.S.I.C./Universitat de Val{\`e}ncia \\
  Edificio Institutos de Paterna, Apartado: 22085, E--46071 Valencia, Spain}
\newcommand{\AddrRIVER}{%
Physics Department, University of California, Riverside, CA 92521, USA }
\newcommand{\AddrLisb}{%
 Departamento de F\'\i sica and CFTP, Instituto Superior T\'ecnico\\
          Avenida Rovisco Pais 1, $\:\:$ 1049-001 Lisboa, Portugal }
\begin{document}

\preprint{IFIC/06-21}
\preprint{UCRHEP-T414}

\vspace*{2cm} \title{Minimal supergravity radiative effects on the
  tri-bimaximal neutrino mixing pattern }

\author{M.~Hirsch} \email{mahirsch@ific.uv.es}\affiliation{\AddrAHEP}
\author{Ernest Ma}\email{ma@phyun8.ucr.edu}\affiliation{\AddrRIVER}
\author{J.~C.~Romao}
\email{jorge.romao@ist.utl.pt}\affiliation{\AddrLisb}
\author{J.~W.~F.~Valle} \email{valle@ific.uv.es}
\affiliation{\AddrAHEP} \author{A.~Villanova del Moral}
\email{Albert.Villanova@ific.uv.es}\affiliation{\AddrAHEP}


\begin{abstract}

  We study the stability of the Harrison-Perkins-Scott (HPS) mixing
  pattern, assumed to hold at some high energy scale, against
  supersymmetric radiative corrections. We work in the framework of a
  reference minimal supergravity model (mSUGRA) where supersymmetry
  breaking is universal and flavor-blind at unification.  The
  radiative corrections considered include both RGE running as well as
  threshold effects.  We find that in this case the solar mixing angle
  can only increase with respect to the HPS reference value, while the
  atmospheric and reactor mixing angles remain essentially stable.
  Deviations from the solar angle HPS prediction towards lower values
  would signal novel contributions from physics beyond the simplest
  mSUGRA model.

\end{abstract}



\maketitle

\section{Introduction
\label{Introduction}}

The discovery of neutrino
oscillations~\cite{fukuda:2002pe,ahmad:2002jz,araki:2004mb,Kajita:2004ga,ahn:2002up}
has indicated a very peculiar structure of lepton
mixing~\cite{Maltoni:2004ei}, quite distinct from that of quarks.
These data have triggered a rush of papers attempting to understand the
values of the leptonic mixing angles from underlying symmetries at a
fundamental level.
An attractive possibility is that the observed pattern of lepton
mixing results from some kind of flavour symmetry, such as $A_4$,
valid at a some superhigh energy scale where the dimension-five
neutrino mass operator arises~\cite{babu:2002dz}.

Here we reconsider the Harrison-Perkins-Scott (HPS) mixing
pattern~\cite{Harrison:2002er} within a simple reference model
approach.  Our only assumption is that at the high energy scale the
tree-level neutrino mass matrix $m_{\nu}^{\textrm{tree}}$ is
diagonalized by the so-called HPS matrix, taken as,
\begin{equation}
\label{eq:HPS}
U_{\textrm{HPS}} = 
\begin{pmatrix}
\sqrt{2/3} & 1/\sqrt{3} & 0\\
-1/\sqrt{6} & 1/\sqrt{3} & -1/\sqrt{2}\\
-1/\sqrt{6} & 1/\sqrt{3} & 1/\sqrt{2}
\end{pmatrix}\;,
\end{equation}
which corresponds to the following mixing angle values:
\begin{align}
\begin{split}
\tan^2\theta_{\Atm}&=\tan^2\theta_{23}^0=1\,,\\
\sin^2\theta_{\textrm{Chooz}}&=\sin^2\theta_{13}^0=0\,,\\
\tan^2\theta_{\Sol}&=\tan^2\theta_{12}^0=0.5\;.
\end{split}
\end{align}
These predictions which hold at high energies may be regarded as a
good first approximation to the observed values~\cite{Maltoni:2004ei}
indicated by oscillation
experiments~\cite{fukuda:2002pe,ahmad:2002jz,araki:2004mb,Kajita:2004ga,ahn:2002up}.
The diagonal neutrino mass matrix can be written as
$ \hat{m}_{\nu}^{\textrm{tree}}=U_{\textrm{HPS}}^{\textrm{T}}\cdot m_{\nu}^{\textrm{tree}}\cdot  U_{\textrm{HPS}}=
{\rm diag} (m_1, m_2, m_3)$,
so that the tree-level neutrino mass matrix becomes
\begin{equation}
\label{eq:m-neu-anz}
m_{\nu}^{\textrm{tree}} = 
\begin{pmatrix}
\frac{2}{3}m_1+\frac{1}{3}m_2 & -\frac{1}{3}m_1+\frac{1}{3}m_2 & -\frac{1}{3}m_1+\frac{1}{3}m_2\\[+1mm]
-\frac{1}{3}m_1+\frac{1}{3}m_2 & \frac{1}{6}m_1+\frac{1}{3}m_2+\frac{1}{2}m_3 & \frac{1}{6}m_1+\frac{1}{3}m_2-\frac{1}{2}m_3\\[+1mm]
-\frac{1}{3}m_1+\frac{1}{3}m_2 & \frac{1}{6}m_1+\frac{1}{3}m_2-\frac{1}{2}m_3 & \frac{1}{6}m_1+\frac{1}{3}m_2+\frac{1}{2}m_3\\
\end{pmatrix}\;.
\end{equation}
This form corresponds to a specific structure for the dimension-five
lepton number violating operator. 
\begin{figure}[h] \centering
    \includegraphics[height=3.5cm,width=.4\linewidth]{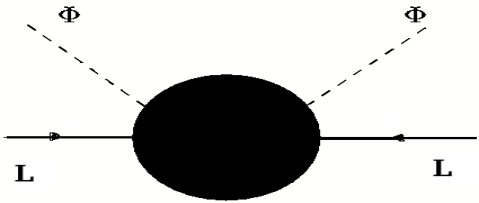}
    \caption{\label{fig:d-5} Dimension five operator responsible for
      neutrino mass.}
\end{figure}
For example, it constitutes the most general ansatz that follows from
a basic $A_4$ symmetry for the neutrino mass matrix and the quark
mixing matrix~\cite{babu:2002dz}. One of the central open questions in
neutrino physics is to identify the exact mechanism of producing
Fig.~\ref{fig:d-5}. As a first step, here we will adopt a
model-independent approach of considering the implications of
Eq.~(\ref{eq:m-neu-anz}) assuming only the evolution expected in
flavor-blind softly broken minimal supergravity at unification. This
will provide us with a reference value that can be useful in the
future for treating different models of neutrino
mass~\cite{Valle:2006vb}.

\section{Radiative corrections
\label{corrections}}

\begin{figure}
  \centering
\includegraphics[height=8cm,width=.8\textwidth]{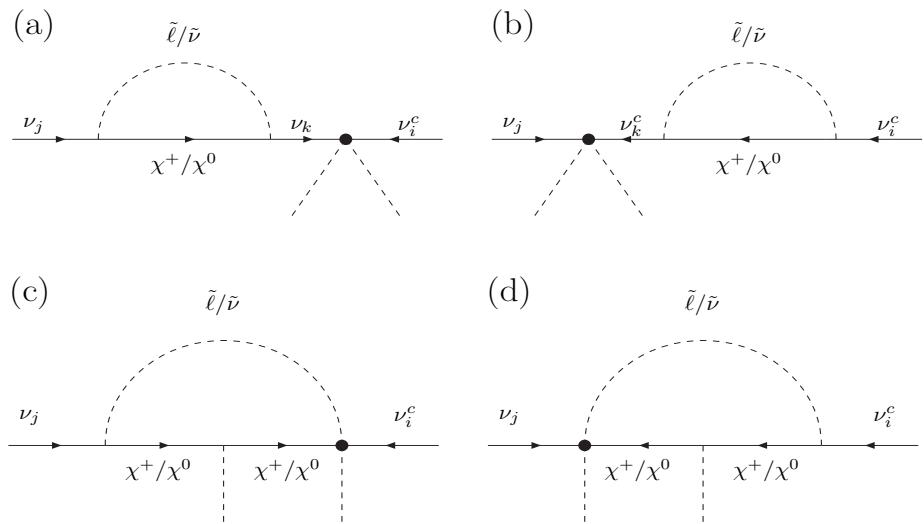}
\caption{Feynman diagrams responsible for neutrino mass radiative
  corrections. The blob indicates an effective Lagrangian term
  obtained from integrating out the heavy right-handed neutrinos}
\label{loopfig}
\end{figure}

It has already been noted that radiative corrections present in the
Standard Model renormalization group equations (RGEs), leave the HPS
``reference'' predictions essentially stable~\cite{Luo:2005fc}.  In
addition to Minimal Supersymmetric Standard Model RGE evolution, here
we consider also the effect of one-loop threshold
effects~\cite{Chun:1999vb}.  
We will first consider the evolution of the neutrino oscillation
parameters that follow from Eq.~(\ref{eq:m-neu-anz}), which covers
both the cases of degenerate as well as
hierarchical neutrino masses. The radiatively corrected neutrino mass
matrix in this case becomes
\begin{equation}\label{eq:m-neu-1loop}
m_{\nu}^{\textrm{1-loop}}=m_{\nu}^{\textrm{tree}}+\hat{\delta}^{\textrm{T}}\cdot m_{\nu}^{\textrm{tree}}+m_{\nu}^{\textrm{tree}}\cdot\hat{\delta}\;,
\end{equation}
where
\begin{equation}
\label{eq:delta-matrix}
\hat{\delta} = 
\begin{pmatrix}
\delta_{ee}' & \delta_{\mu e} & \delta_{\tau e}\\
\delta_{e\mu} & \delta_{\mu\mu}' & \delta_{\tau\mu}\\
\delta_{e\tau} & \delta_{\mu\tau} & \delta_{\tau\tau}'\\
\end{pmatrix}\;.
\end{equation}
The diagonal elements include the threshold correction and the RGE running
\begin{equation}
\label{eq:delta-diag}
\delta_{\alpha\alpha}'=\delta_{\alpha\alpha}+\delta_{\alpha}\;,
\end{equation}
where the RGE running effect is~\cite{Babu:1993qv}
\begin{equation}\label{eq:del-RGE}
\delta_{\alpha} = \frac{-h_{\alpha}^2}{16\pi^2}\ln\left(\frac{M_{\textrm{GUT}}}{M_{\textrm{EWSB}}}\right)\;.
\end{equation}
In order to get the analytic expressions for the threshold
corrections, we proceed as in Ref.~\cite{Hirsch:2003dr}. However, now
we do not neglect Yukawa couplings, taking into account the fact that
right- and left-handed charged sleptons mix. Therefore, the analytic
expressions for the deltas are
\begin{equation}
\begin{split}
\label{radneumass}
\delta_{\alpha \beta}^{{\rm (a)}\chi^+} &= 
\sum_{i=1}^6 \sum_{A=1}^2\frac{1}{16\pi^2}
(g U_{A1}^*R_{i\alpha}^{\tilde\ell}-h_{\alpha}U_{A2}^*R_{i\alpha+3}^{\tilde\ell})
(g U_{A1}R_{i\beta}^{\tilde\ell\ast}-h_{\beta}U_{A2}R_{i\beta+3}^{\tilde\ell\ast})\\
&\quad\times B_1(m_{\chi_{A}^+}^2,m_{\tilde\ell_i}^2)  \;,  \\
\delta_{\alpha \beta}^{{\rm (a)}\chi^0} &= 
\sum_{i=1}^3 \sum_{A=1}^4 \frac{1}{32\pi^2}|g N_{A2}-g' N_{A1}|^2
R_{i\alpha}^{\tilde\nu} R_{i \beta}^{\tilde\nu\ast}  
B_1(m_{\chi_{A}^0}^2,m_{\tilde\nu_i}^2) \;, \\
\delta_{\alpha \beta}^{{\rm (c)}\chi^+} &= 
\sum_{i=1}^6 \sum_{A=1}^2 \sum_{B=1}^2\frac{1}{4\pi^2} 
(g U_{A1}^*R_{i\alpha}^{\tilde\ell}-h_{\alpha}U_{A2}^*R_{i\alpha+3}^{\tilde\ell})
g U_{A1}|V_{B2}|^2 R_{i \beta}^{\tilde\ell\ast}
C_{00}(m_{\chi_{A}^+}^2,
m_{\chi_{B}^+}^2 ,m_{\tilde\ell_i}^2) \;,  \\
\delta_{\alpha \beta}^{{\rm (c)}\chi^0} &=
\sum_{i=1}^3 \sum_{A=1}^4 \sum_{B=1}^4 \frac{1}{8\pi^2}
|gN_{A2}-g'N_{A1}|^2|N_{B4}|^2 
R_{i\alpha}^{\tilde\nu} R_{i \beta}^{\tilde\nu\ast} 
C_{00}(m_{\chi_{A}^0}^2,
m_{\chi_{B}^0}^2,m_{\tilde\nu_i}^2) \;, 
\end{split}
\end{equation}
where we have evaluated the Feynman diagrams at zero external
momentum, which is an excellent approximation as the neutrino masses
are tiny.  Here $\delta_{\alpha \beta}^{{\rm (a,c)}\chi^+},
(\alpha,\beta=e,\mu,\tau$), are the contributions from the
chargino/charged slepton diagrams in Fig.~\ref{loopfig} (a,c), respectively, 
while $\delta_{\alpha \beta}^{{\rm (a,c)}\chi^0}$ are the contributions 
from the neutralino/sneutrino diagrams.
The values of the $\delta_{\alpha\beta}$'s, in Eqs. (\ref{eq:delta-matrix})
and (\ref{eq:delta-diag}) 
are the sum of the four contributions given above.
Analogous contributions exist corresponding to the symmetrized terms
in Eq.~(\ref{eq:m-neu-1loop}), required by the Pauli principle, as
displayed in Fig.~\ref{loopfig} (b,d).  
In the above formulas, $U$ and $V$ are the chargino mixing matrices
and $m_{\chi^+_A}, (A=1,2)$, are chargino masses, while $N$ is the
neutralino mixing matrix with $m_{\chi^0_A}, (A=1,..,4)$, denoting the
neutralino masses.  Finally, the matrices $R^{\tilde\ell/\tilde\nu}$
denote the slepton/sneutrino mixing matrices, respectively. The
coupling constant of the $SU(2)$ gauge group is denoted $g$ and that
of $U(1)$ is $g'$. Here $h_{\alpha}$ is the charged lepton Yukawa
coupling in the basis where the charged lepton masses are diagonal.
Furthermore $B_1$ and $C_{00}$ are Passarino-Veltman functions given
by
\begin{equation}
 B_1(m_0^2,m_1^2)=-\frac{1}{2}\Delta_\epsilon +\frac{1}{2} 
\ln \left( \frac{m_0^2}{M^2_{\textrm{EWSB}}} \right)+
\frac{-3+4t-t^2-4t\ln (t)+2t^2\ln(t)}{4(t-1)^2}\;,
\end{equation}
where $t=m_1^2/m_0^2$ and
\begin{equation}
C_{00}(m_0^2,m_1^2,m_2^2) = 
\frac{1}{8}(3+2\Delta_{\epsilon}) -\frac{1}{4} \ln \left( \frac{m_0^2}{M^2_{\textrm{EWSB}}} \right)
 +  \frac{-2r_1^2(r_2-1)\ln(r_1)+2r_2^2(r_1-1)\ln(r_2)}
{8(r_1-1)(r_2-1)(r_1-r_2)}\;,
\end{equation}
where $r_1=m_1^2/m_0^2$ and $r_2=m_2^2/m_0^2$. We have 
used dimensional regularization, with  $\epsilon=4-n$ and $n$ is 
the number of space-time dimensions. The term $\Delta_{\epsilon}=(2/\epsilon)-\gamma+4\ln(4\pi)$, where $\gamma$ is Euler's constant, is divergent as $\epsilon\to 0$.

\section{Corrections to mixing angles: numerical results}
\label{sec:corr-neutr-mixing}

We now describe our numerical procedure.  In order to compute the
magnitude of the radiative corrections expected in the HPS anzatz we
work in the framework of a reference minimal supergravity model
approach, with universal flavor-blind soft supersymmetry breaking
terms at unification. Therefore the off-diagonal elements in the
matrix in Eq.~(\ref{eq:delta-matrix}) are all zero~\footnote{Nonzero
  off-diagonal elements may arise due to running, see discussion.}
\begin{equation}
\label{eq:zero-off-diag}
\delta_{e\mu}=\delta_{e\tau}=\delta_{\mu\tau}=
\delta_{\mu e}=\delta_{\tau e}=\delta_{\tau \mu}=0\;.
\end{equation}

We first have used the SPheno package~\cite{Porod:2003um} to calculate
spectra and mixing matrices within mSUGRA within the ranges:
$M_{1/2},\,M_0,\,A_0\in[100,\,1000]$ GeV, $A_0$ with both signs,
$\tan\beta\in[2.5,\,50]$ and $\mu$ with both signs.  Then we have
calculated the RGE running, Eq.~(\ref{eq:del-RGE}), and the threshold
corrections, Eqs.~(\ref{radneumass}).  We have explicitly checked that
the dominant contribution to $\delta'_{\alpha\alpha}$, defined in
Eq.~(\ref{eq:delta-diag}), always comes from the threshold corrections
for $\alpha=e,\,\mu$.  
Also for $\alpha=\tau$, threshold corrections are usually more important 
than RGE running contributions,
typically
\begin{equation}
\delta_{\alpha\alpha}\sim\mathcal{O}(10^{(-4,-3)})\,,\quad \forall\alpha
\end{equation}
while
\begin{align}
\delta_{e}&\sim\mathcal{O}(10^{(-11,-9)}) & \delta_{\mu}&\sim\mathcal{O}(10^{(-7,-4)}) & \delta_{\tau}&\sim\mathcal{O}(10^{(-4,-2)}) \,. 
\end{align}
Note that only for very large values of $\tan\beta$, the RGE effect 
$\delta_{\tau}$ is slightly larger than the threshold corrrections 
$\delta_{\tau\tau}$.
Using these radiative corrections we have computed the delta matrix in
Eq.~(\ref{eq:delta-matrix}) and inserted it in the neutrino mass
matrix at 1-loop given in Eq.~(\ref{eq:m-neu-1loop}).  We have then
numerically diagonalized the 1-loop neutrino mass matrix in
Eq.~(\ref{eq:m-neu-1loop}) in order to obtain the neutrino masses and
mixing angles.

Notice that the HPS scheme only fixes neutrino mixing angles.  Thus,
the neutrino masses are free parameters.  Making use of this freedom,
we have used an iterative procedure in order to choose the parameters
$m_1$, $m_2$ and $m_3$, so that the numerically calculated 1-loop
neutrino masses are such that the solar and atmospheric squared-mass
splittings $\Dms$ and $\Dma$ reproduce the current best fit point
value. In our numerical calculation we concentrate on normal hierarchy. 
We will comment on the case of inverse hierarchy at the end of the next section.

The numerically calculated atmospheric and reactor neutrino mixing
angles at low energies do not deviate significantly from its HPS
reference value at high energies.  Indeed, the numerical results are:
\begin{align}
\begin{split}
\tan^2\theta_{\Atm}&\lesssim\tan^2\theta_{23}^0+\mathcal{O}(10^{-1})\,,\\
\sin^2\theta_{\textrm{Chooz}}&\lesssim\sin^2\theta_{13}^0+\mathcal{O}(10^{-7})\;.
\end{split}
\end{align}

\begin{figure}[htbp]
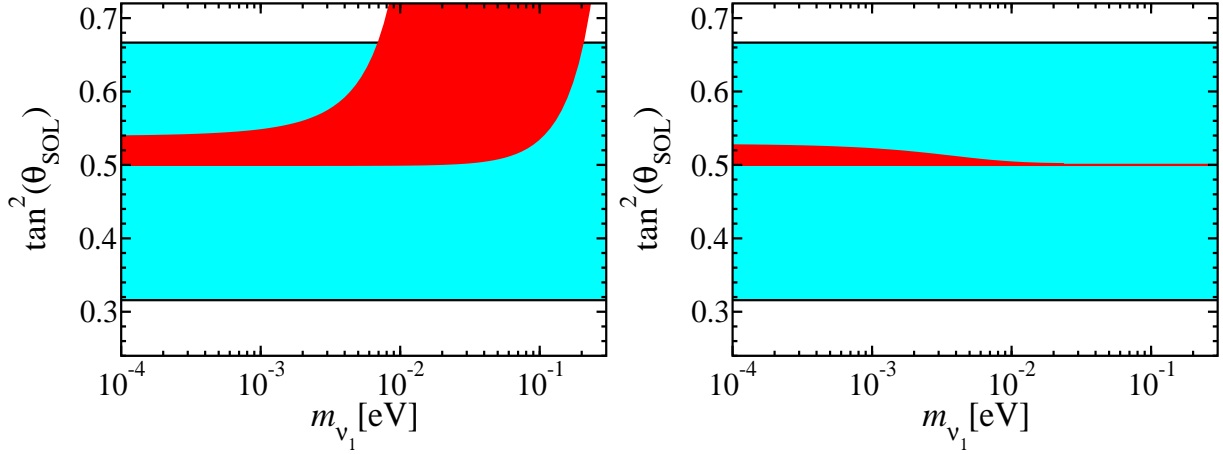

  \centering
\includegraphics[height=6cm,width=8cm]{tansqsol-mnu1-CP-pp.eps}
\includegraphics[clip,height=6cm,width=8cm]{tansqsol-mnu1-CP-pm.eps}
\caption{Upper bound for the solar mixing parameter
  $\tan^2\theta_{\Sol}$, as a function of $m_{\nu_1}$ (in eV), for
  $\tan\beta=2.5$ (lower border of the red band) and $\tan\beta=50$
  (upper border of the red band).  On the left panel, $m_{\nu_1}$ and
  $m_{\nu_2}$ have the same $CP$ sign.  On the right panel,
  $m_{\nu_1}$ and $m_{\nu_2}$ have opposite $CP$ sign.  The neutrino
  mass splittings are assumed to have their best fit value
  from~\cite{Maltoni:2004ei}.  The horizontal band corresponds to the
  $3\sigma$ allowed range for
  $\tan^2\theta_{\Sol}$~\cite{Maltoni:2004ei}.}
  \label{fig:tansqsol-mnu1}
\end{figure}

However, the solar neutrino mixing angle can be significantly
affected.  In Fig.~\ref{fig:tansqsol-mnu1}, we have plotted the
maximum deviation of the solar angle from the HPS reference value for
$\tan\beta\in[2.5,50]$, as a function of $m_{\nu_1}$, for both extreme
$CP$ parity combinations for $m_{\nu_1}$ and $m_{\nu_2}$: same sign
(left panel) and opposite sign (right panel).  All the other $CP$
possiblities lie in between these two extreme cases.
As can be seen, the solar mixing angle remains essentially stable in
the case of opposite $CP$ signs, while deviations are maximal in the
case of same $CP$ signs. In this case, the solar mixing angle always
increases with respect to the HPS value, irrespective of mSUGRA
parameters. 
Moreover we can get a rough upper bound on $m_{\nu_1}$ of order
\begin{equation}
\label{eq:bound}
m_{\nu_1}\lsim 0.2 \;\textrm{eV}
\end{equation}
for the mSUGRA parameter values: $M_{1/2} = 100$ GeV, $M_0 = -
A_0=10^3$ GeV, $\mu >0$ and $\tan\beta=2.5$.
Note that the upper bound is sensitive to the values of $\tan\beta$.
For higher values of $\tan\beta$ the radiative corrections are larger,
implying a more stringent bound on $m_{\nu_1}$, as indicated by the
upper boundary of the red (dark) band of the left panel in
Fig.~\ref{fig:tansqsol-mnu1}.
Here we have fixed solar and atmospheric mass squared splittings at
their best-fit values from Ref.~\cite{Maltoni:2004ei}.  However, we
have explicitly checked that the effect of letting $\Dma$ and $\Dms$
vary within their current $3\sigma$ allowed range is negligible, i.~e.
the bands obatined at the extreme values almost coincide with the ones
in Fig.~\ref{fig:tansqsol-mnu1}.
%


\section{Analytical understanding}
\label{sec:analyt-underst}

The numerical results presented above can be understood analytically
as follows.  If we perform the original HPS rotation to the 1-loop
neutrino mass matrix in Eq.~\ref{eq:m-neu-1loop}, we get:
\begin{align}
\hat{m}_{\nu}^{\textrm{1-loop}}&=U_{\textrm{HPS}}^{\textrm{T}}\cdot m_{\nu}^{\textrm{1-loop}}\cdot U_{\textrm{HPS}}\\[+1mm]
\label{eq:diag1loop}
&=
\begin{pmatrix}
(1+\delta_{11})m_1 & \delta_{12}^{m_1}m_1+\delta_{12}^{m_2}m_2 & \delta_{13}^{m_1}m_1+\delta_{13}^{m_3}m_3\\[+2mm]
\delta_{12}^{m_1}m_1+\delta_{12}^{m_2}m_2 & (1+\delta_{22})m_2 & \delta_{23}^{m_2}m_2+\delta_{23}^{m_3}m_3\\[+2mm]
\delta_{13}^{m_1}m_1+\delta_{13}^{m_3}m_3 & \delta_{23}^{m_2}m_2+\delta_{23}^{m_3}m_3 & (1+\delta_{33})m_3
\end{pmatrix}\;,
\end{align}
where
\begin{eqnarray}
\label{eq:d11}
\delta_{11} & = &\frac{1}{3}(4 \delta_{ee}' + \delta_{\mu\mu}' +
\delta_{\tau\tau}' - 2 \delta_{e\mu} - 2 \delta_{\mu e} - 2
\delta_{e\tau} - 2 \delta_{\tau e} + \delta_{\mu\tau} +
\delta_{\tau\mu})\;,  \nonumber \\
\delta_{22} & = &\frac{2}{3} ( \delta_{ee}' + \delta_{\mu\mu}' + \delta_{\tau\tau}' + \delta_{e\mu} + \delta_{\mu e} + \delta_{e\tau} + \delta_{\tau e} + \delta_{\mu\tau} + \delta_{\tau\mu} )\;,  \nonumber\\
\delta_{33} & = &\delta_{\mu\mu}' + \delta_{\tau\tau}' - \delta_{\mu\tau} - \delta_{\tau\mu} 
\;, \nonumber\\
\delta_{12}^{m_1} & =& \frac{1}{3\sqrt{2}}(2 \delta_{ee}' - \delta_{\mu\mu}' - \delta_{\tau\tau}' - \delta_{e\mu} + 2 \delta_{\mu e} - \delta_{e\tau} + 2 \delta_{\tau e} - \delta_{\mu\tau} - \delta_{\tau\mu}) \;, \nonumber\\
\delta_{12}^{m_2} & = &\frac{1}{3\sqrt{2}}(2 \delta_{ee}' -
\delta_{\mu\mu}' - \delta_{\tau\tau}' + 2 \delta_{e\mu} - \delta_{\mu
  e} + 2 \delta_{e\tau} - \delta_{\tau e} - \delta_{\mu\tau} -
\delta_{\tau\mu})\;,  \\
\delta_{13}^{m_1} & = &\frac{1}{2\sqrt{3}}(\delta_{\mu\mu}' -
\delta_{\tau\tau}' - 2 \delta_{\mu e} + 2 \delta_{\tau e} +
\delta_{\mu\tau} - \delta_{\tau\mu}) \;, \nonumber \\
\delta_{13}^{m_3} & = &\frac{1}{2\sqrt{3}}(\delta_{\mu\mu}' -
\delta_{\tau\tau}' - 2 \delta_{e\mu} + 2 \delta_{e\tau} -
\delta_{\mu\tau} + \delta_{\tau\mu})\;,  \nonumber \\
\delta_{23}^{m_2} & = &\frac{1}{\sqrt{6}}(-\delta_{\mu\mu}' +
\delta_{\tau\tau}' - \delta_{\mu e} + \delta_{\tau e} -
\delta_{\mu\tau} + \delta_{\tau\mu}) \;, \nonumber \\
\delta_{23}^{m_3} & = &\frac{1}{\sqrt{6}}(-\delta_{\mu\mu}' + \delta_{\tau\tau}' - \delta_{e\mu} + \delta_{e\tau} + \delta_{\mu\tau} - \delta_{\tau\mu}) \nonumber\;.
\end{eqnarray}
The matrix in Eq.~(\ref{eq:diag1loop}) should be nearly diagonal and
its off-diagonal elements determine the deviations from
tri-bimaximality.  We define the following parameters characterizing
the deviations from tri-bimaximality:
\begin{equation}
\label{eq:epsij}
\epsilon_{ij}\simeq\frac{1}{2}\tan(2\epsilon_{ij})=\frac{(\hat{m}_{\nu}^{\textrm{1-loop}})_{ij}}
{(\hat{m}_{\nu}^{\textrm{1-loop}})_{jj}-(\hat{m}_{\nu}^{\textrm{1-loop}})_{ii}}\;,
\end{equation}
so that
\begin{align}
\begin{split}\label{eq:angles}
\theta_{\Atm}&\equiv\theta_{23}\simeq\theta_{23}^0+\epsilon_{23}\;,\\
\theta_{\textrm{Chooz}}&\equiv\theta_{13}\simeq\theta_{13}^0+\epsilon_{13}\;, \\
\theta_{\Sol}&\equiv\theta_{12}\simeq\theta_{12}^0+\epsilon_{12}\;. 
\end{split}
\end{align}
Substituting the matrix elements in Eq.~(\ref{eq:diag1loop}) into Eq.~(\ref{eq:epsij}), we get:
\begin{align}
\label{eq:eps23}
\epsilon_{23} & = \frac{\delta_{23}^{m_2} m_2 + \delta_{23}^{m_3} m_3}{(-1 - \delta_{22}) m_2 + (1 + \delta_{33}) m_3}\;, \\
\epsilon_{13} & = \frac{\delta_{13}^{m_1} m_1 + \delta_{13}^{m_3} m_3}{(-1 - \delta_{11}) m_1 + (1 + \delta_{33}) m_3}\;, \\
\label{eq:eps12}
\epsilon_{12} & = \frac{\delta_{12}^{m_1} m_1 + \delta_{12}^{m_2} m_2}{(-1 - \delta_{11}) m_1 + (1 + \delta_{22}) m_2} \;.
\end{align}
Taking into account that for mSUGRA the off-diagonal elements in the
matrix in Eq.~(\ref{eq:delta-matrix}) are all zero, see
Eq.(\ref{eq:zero-off-diag}), the $\delta$'s in Eq.~(\ref{eq:d11})
become
\begin{equation}
\begin{split}
\label{eq:d11-mSUGRA}
\delta_{11} & = \delta_{11}^0 = \frac{1}{3}(4 \delta_{ee}' + \delta_{\mu\mu}' + \delta_{\tau\tau}' ) \;,\\ 
\delta_{22} & = \delta_{22}^0 = \frac{2}{3} ( \delta_{ee}' + \delta_{\mu\mu}' + \delta_{\tau\tau}' ) \;,\\ 
\delta_{33} & = \delta_{33}^0 = \delta_{\mu\mu}' + \delta_{\tau\tau}' \;,\\ 
\delta_{12}^{m_1} & = \delta_{12}^{m_2} = \delta_{12}^0 = \frac{1}{3\sqrt{2}}(2 \delta_{ee}' - \delta_{\mu\mu}' - \delta_{\tau\tau}' ) \;,\\ 
\delta_{13}^{m_1} & = \delta_{13}^{m_3} =\delta_{13}^0 = \frac{1}{2\sqrt{3}}(\delta_{\mu\mu}' - \delta_{\tau\tau}' ) \;,
\\ 
\delta_{23}^{m_2} & = \delta_{23}^{m_3} =\delta_{23}^0 = \frac{-1}{\sqrt{6}}(\delta_{\mu\mu}' - \delta_{\tau\tau}' ) \;.
\end{split}
\end{equation}
The deviations of the neutrino mixing angles from the HPS value given
in Eqs.~(\ref{eq:eps23}-\ref{eq:eps12}) then become
\begin{align}
\label{eq:eps23-mSUGRA}
\epsilon_{23} & = \frac{\delta_{23}^0 ( m_2 + m_3 )}{(-1 - \delta_{22}^0) m_2 + (1 + \delta_{33}^0) m_3}\;, \\
\label{eq:eps13-mSUGRA}
\epsilon_{13} & = \frac{\delta_{13}^0 ( m_1 + m_3 )}{(-1 - \delta_{11}^0) m_1 + (1 + \delta_{33}^0) m_3}\;, \\
\label{eq:eps12-mSUGRA}
\epsilon_{12} & = \frac{\delta_{12}^0 ( m_1 + m_2 )}{(-1 - \delta_{11}^0) m_1 + (1 + \delta_{22}^0) m_2} \;.
\end{align}
If $\epsilon_{12}$, given in  Eq.~(\ref{eq:eps12-mSUGRA}), 
is always positive, $\theta_{\Sol}$ can only increase, 
see Eq.~(\ref{eq:angles}). The denominator in
Eq.~(\ref{eq:eps12-mSUGRA}) can be approximated to
\begin{equation}
(-1 - \delta_{11}^0) m_1 + (1 + \delta_{22}^0)) m_2 \simeq - m_1 + m_2 > 0
\end{equation}
and hence, by assumption, is always positive. The sign of
$\epsilon_{12}$ will be the sign of $\delta_{12}^0$ given by
Eq.~(\ref{eq:d11-mSUGRA}).  Considering the expressions for the deltas
given in Eq.~(\ref{radneumass}) and bearing in mind that the
Passarino-Veltmann functions depend rather smoothly on their
arguments, we can take them out of the sum.  The following very rough
estimations of the threshold corrections result
\begin{align}\begin{split}\label{eq:del-ap-1}
\delta_{\alpha\alpha}&\simeq\frac{1}{32\pi^2}(3g^2(B_1+4C_{00})+g'^2(B_1+4C_{00}))\;,\qquad (\alpha=e,\mu)\;,\\
\delta_{\tau\tau}&\simeq\frac{1}{32\pi^2}(3g^2(B_1+4C_{00})+g'^2(B_1+4C_{00})+2h^{2}_{\tau}B_1)\;,
\end{split}\end{align}
where we have neglected the charged lepton Yukawa couplings for
$\alpha=e,\mu$. Using
\begin{equation}
\lim_{m_{\tilde L_i}^2\to\infty}\frac{B_1(m_{\chi_{A}}^2,m_{\tilde L_i}^2)}{C_{00}(m_{\chi_{A}}^2,m_{\chi_{B}}^2,m_{\tilde L_i}^2)}
=-2\;,
\end{equation}
Eq.~(\ref{eq:del-ap-1}) becomes
\begin{align}\begin{split}\label{eq:del-ap-2}
\delta_{\alpha\alpha}&\simeq\frac{-B_1}{32\pi^2}(3g^2+g'^2)\;,\qquad (\alpha=e,\mu)\;,\\
\delta_{\tau\tau}&\simeq\frac{-B_1}{32\pi^2}(3g^2+g'^2-2h^{2}_{\tau})\;.
\end{split}\end{align}
Therefore, the  contribution of the threshold corrections to $\delta_{12}^0$ is roughly
\begin{equation}
2\delta_{ee}-\delta_{\mu\mu}-\delta_{\tau\tau}\simeq\frac{-B_1}{16\pi^2}h_{\tau}^2\;.
\end{equation}
Besides the threshold correction contributions, one has also to
consider the RGE running contribution.  Here the dominant part
obviously is $\delta_{\tau}$, given in Eq.~(\ref{eq:del-RGE}). The
approximated expression for $\delta_{12}^0$, defined in
Eq.~(\ref{eq:eps12-mSUGRA}), is then
\begin{equation}\label{eq:apr-d12-1}
\delta_{12}^0\simeq \frac{1}{3\sqrt2} \left( 2\delta_{ee}-\delta_{\mu\mu}-\delta_{\tau\tau}-\delta_{\tau} \right)
\simeq \frac{1}{3\sqrt2} \frac{h_{\tau}^2}{16\pi^2}\left[-B_1+\ln\left(\frac{M_{\textrm{GUT}}}{M_{\textrm{EWSB}}}\right) \right]\;.
\end{equation}
Considering that in the limit where the slepton mass goes to infinity,
the Passarino-Veltman function $B_1$ behaves as
\begin{equation}
\lim_{m_{\tilde L_i}^2\to\infty}B_1(m_{\chi_{A}}^2,m_{\tilde L_i}^2)
\simeq\frac{1}{2}\ln\left(\frac{m_{\tilde L_i}^2}{m_{\chi_{A}}^2}\right)\;,
\end{equation}
one obtains, from Eq.~(\ref{eq:apr-d12-1}),
\begin{equation}\label{eq:apr-d12-2}
\delta_{12}^0\simeq \frac{1}{3\sqrt2} \left(2\delta_{ee}-\delta_{\mu\mu}-\delta_{\tau\tau}-\delta_{\tau} \right)
\simeq \frac{1}{3\sqrt2} \frac{h_{\tau}^2}{16\pi^2}\left[\ln\left(\frac{M_{\textrm{GUT}}}{M_{\textrm{EWSB}}}\right)
-\ln\left(\frac{m_{\tilde L_i}}{m_{\chi_{A}}}\right)\right]\;,
\end{equation}
which is always positive, thus explaining why $\epsilon_{12}>0$.  
Note that although the threshold corrections are in general larger
than the RGE contributions, in $\delta_{12}^0$ there is a cancellation
among the threshold corrections so that the $\delta_{\tau}$ RGE
contribution becomes the relevant term. We have numerically checked
that
\begin{equation}
2\delta_{ee}-\delta_{\mu\mu}-\delta_{\tau\tau}\sim\mathcal{O}(10^{(-6,-3)})\;.
\end{equation}
This cancellation among the threshold corrections is the reason why
the solar neutrino mixing angle can only increase with respect its HPS
reference value.

We now turn to the other two neutrino mixing angles.  In the mSUGRA
framework the deviations from the HPS predictions are much smaller
than found for the solar mixing parameter, and fit within their
current experimental $3\sigma$ allowed range given in
Ref.~\cite{Maltoni:2004ei} for acceptable $m_{\nu_1}$ values.
The reason for this can be understood from
Eqs.~(\ref{eq:eps23-mSUGRA}-\ref{eq:eps12-mSUGRA}). On the one hand,
the deltas on the numerators, given by Eq.~(\ref{eq:d11-mSUGRA}), are
all of the same order.  For small values of $m_{\nu_1}$ the deviations
are all negligible, since they are all proportional to the previous
deltas. For large $m_{\nu_1}$ values the neutrino masses are nearly
degenerate so that the numerators in
Eqs.~(\ref{eq:eps23-mSUGRA}-\ref{eq:eps12-mSUGRA}) are all of the same
order.  The denominators in
Eqs.~(\ref{eq:eps23-mSUGRA}-\ref{eq:eps12-mSUGRA}) can be approximated
as
\begin{align}
\label{eq:denominators-1}
(-1 - \delta_{22}^0) m_2 + (1 + \delta_{33}^0) m_3 & \simeq m_3 - m_2\;,\\
(-1 - \delta_{11}^0) m_1 + (1 + \delta_{33}^0) m_3 & \simeq  m_3 - m_1\;,\\
(-1 - \delta_{11}^0) m_1 + (1 + \delta_{22}^0) m_2 & \simeq  m_2 - m_1\;.
\label{eq:denominators-2}
\end{align}
Although these mass differences are very small, $m_3-m_2$ and
$m_3-m_1$ are larger than $m_2-m_1$, thus making $\epsilon_{23}$ and
$\epsilon_{13}$ smaller than $\epsilon_{12}$. 

We now comment briefly on inverse hierarchy. As can be seen from Eqs.~(\ref{eq:denominators-1}-\ref{eq:denominators-2}), 
for inverse hierarchy, $m_2-m_1$ is still much smaller than $m_3-m_2$ or $m_3-m_1$, while the latter two 
just change sign but not the magnitude. We therefore expect that the above discussion remains essentially correct
also for inverse hierarchy.

We should stress that we have considered so far the $CP$
conserving case HPS ansatz, with same-$CP$-sign neutrino mass
eigenvalues,
\begin{equation}
m_1,m_2,m_3>0\;.
\end{equation}
However, for all other CP combinations the denominators in
Eqs.~(\ref{eq:eps23-mSUGRA}-\ref{eq:eps12-mSUGRA}) are larger such
that the deviations from HPS mixing pattern become smaller and
correspondingly relax the bound in Eq.~(\ref{eq:bound}). In particular
for the case of opposite CP signs there is no bound, as seen in right
panel in Fig.~\ref{fig:tansqsol-mnu1}.


\section{Summary and discussion}

We have studied the stability of the HPS mixing ansatz that could
arise from a flavor symmetry valid at some high energy scale, against
supersymmetric radiative corrections (RGE running and threshold
effects). We have adopted a model-independent minimal supergravity
framework where supersymmetry breaking is universal and flavor-blind
at unification.  In this case we have found that the solar mixing
angle can only be increased with respect to the HPS reference value.
Under the assumption that all neutrino masses have the same $CP$-sign,
this sets a rough upper bound on the mass of the lightest neutrino
which, in turn, implies that the neutrinoless double beta decay rate
is also bounded as a function of the mSUGRA parameters.  In contrast,
in the case of opposite CP signs there is no bound on the lightest
neutrino mass.  We have also shown that the atmospheric and reactor
mixing angles remain essentially stable in all cases.  It should not
be surprising that the effect of radiative corrections is more
important for the solar angle than for the others.  It simply reflects
the fact that the solar is the smallest of the two neutrino mass
splittings.


We stress that in our approach we have assumed only that the matrix
$m_{\nu}^{\textrm{tree}}$ is diagonalized by the HPS matrix at the
unification scale and this gets modified only by minimal supergravity
radiative corrections, universal and flavor-blind at unification.
This concerns the structure of the dimension-five operator,
Fig.~\ref{fig:d-5}. Additional radiative
corrections~\cite{Babu:1993qv} to the solar angle HPS prediction are
expected, if the neutrino mass arises {\sl a la
  seesaw}~\cite{Minkowski:1977sc,Orloff:2005nu,schechter:1980gr,Lazarides:1980nt}.
Their magnitude will be determined by the strength of the Yukawa
coupling characterizing the Dirac neutrino mass entry in the seesaw
mass matrix~\cite{borzumati:1986qx}.
This will depend strongly on the details of the model, in particular,
on whether Higgs triplets are present in the
seesaw~\cite{schechter:1980gr} or on whether the seesaw is
extended~\cite{mohapatra:1986bd}.
Scrutinizing the schemes for which it is possible to decrease the
solar mixing angle value predicted by the HPS mixing pattern towards
its currently preferred best fit point value will be considered
elsewhere~\cite{prep}, together with the related issue of the lepton
flavor violating processes that would be expected in these schemes.

\section*{Acknowledgements}

We thank Werner Porod for useful discussions about SPheno. This work was
supported by Spanish grants FPA2005-01269 and BFM2002-00345 and by the
EC RTN network MRTN-CT-2004-503369. M.~H. was supported by a Ramon y
Cajal contract. E.~M. was supported in part by the U.S. Department of Energy under Grant No. DE-FG03-94ER40837. A.~V.~M. was supported by Generalitat Valenciana.


\renewcommand{\baselinestretch}{1.2}

\end{document}